\documentclass[]{aa}

 \usepackage{graphicx}
 \usepackage{txfonts}

\begin{document}

   \title{Intra-day variability observations and the VLBI structure analysis of
   quasar S4 0917+624}
   \author{X. Liu\inst{1,2}
           \and
           L.-G. Mi\inst{1,3}
           \and
             J. Liu\inst{1,3}
             \and
             L. Cui\inst{1,2}
             \and
             H.-G. Song\inst{1,2}
           \and
         T.P. Krichbaum\inst{4}
         \and
         A. Kraus\inst{4}
         \and
         L. Fuhrmann\inst{4}
         \and
         N. Marchili\inst{5}
         \and
         J.A. Zensus\inst{4}
          }


\institute{Xinjiang Astronomical Observatory, Chinese Academy of
Sciences, 150 Science 1-Street, Urumqi 830011, China\\
\email{liux@xao.ac.cn}\and Key Laboratory of Radio Astronomy,
Chinese Academy of Sciences, 2 West Beijing Road, Nanjing, JiangSu
210008, China \and University of Chinese Academy of Sciences,
Beijing 100049, China \and Max-Plank-Institut f\"ur
Radioastronomie, Auf dem H\"ugel 69, 53121 Bonn, Germany \and
Dipartimento di Fisica e Astronomia, Universit\`a di Padova, via
Marzolo 8 I-35131 Padova, Italy}

    \date{Received / Accepted }

  \abstract
  {}
  {The quasar S4 0917+624 is one of the targets in the Urumqi flux density monitoring program,
   which aims to study the properties of intra-day
   variable (IDV) extragalactic radio sources.}
   {The IDV observations of S4 0917+624 were carried out monthly, from
   August 2005 to January 2010, with the Urumqi 25m radio telescope at 4.8 GHz.
   We analyze these and previous IDV observations
   to investigate the long-term
   IDV characteristics of S4 0917+624.
   We also study the long-term structural variability on milliarcsecond scales using VLBI
   maps obtained at 15 GHz (taken from the MOJAVE database) in order to search for
   a possible relation between variations in the IDV  pattern and the source structure.}
  {The quasar S4 0917+624 exhibits only very weak or no IDV during our 4.5 year observing interval.
Prior to the year 2000, the source S4 0917+624 was one of the most
prominent IDV sources. Our new data indicate that the
previous strong IDV has ceased. We analyzed
the long-term VLBI structural variability using Gaussian
model-fitting. From this we obtained the flux densities and the
deconvolved sizes of core and inner-jet components of the
source. We studied the properties such as core fraction, angular size,
spectral index, and brightness temperature of VLBI core for S4 0917+624, as well as the time delay between
5 and 15 GHz variations, and compared them with the IDV properties of S4 0917+624.
The source shows ejection of several jet components that
are suspected to have partially reduced the IDV amplitude of S4
0917+624. However, during 2005-2006, the VLBI core size was
comparable to the size before the year 2000, but no strong IDV
was detected in the period, suggesting that the quenching effect due to
source size changes may not be responsible for the lack of strong IDV after the year 2000.
The refractive scattering properties for the strong IDV phase of S4 0917+624 before
the year 2000 are discussed.}
  {The disappearance of strong IDV in S4 0917+624 after the year 2000 is a mystery
  and cannot be explained via the quenching effect by changes in the observable VLBI structure.
  However it may be caused by changes in the interstellar medium, i.e.
  by interstellar weather, which induces changes in the scintillation pattern
  on timescales of several years. Further coordinated multi-frequency observations
  will be required to distinguish between the effect of source-intrinsic variability and changing properties of the interstellar medium.}
 { Table 3, which contains all flux measurements from the
Urumqi observations, is only available at the CDS via anonymous
ftp to cdsarc.u-strasbg.fr (130.79.128.5) or via
http://cdsarc.u-strasbg.fr/viz-bin/qcat?J/A+A/.
}


\maketitle

\section{Introduction}
\label{sec:intro} Intra-day variability (IDV) of active galactic
nuclei (AGN) at centimeter wavelengths was discovered in the 1980s
(Witzel et al. 1986; Heeschen et al. 1987). Significant IDV mainly
occurs in flat spectrum radio sources, which belong to the class
of radio-loud core-dominated AGN. Statistical studies suggest an
IDV rate of 25\%-50\% in these flat-spectrum radio sources
(Quirrenbach et al. 1992; Lovell et al. 2008) and an even higher
rate of $\sim$60\% in gamma-ray bright Fermi blazars (Liu et al.
2012a). From the beginning of the discovery of IDV, both
source-intrinsic and -extrinsic mechanisms were proposed to
explain the IDV (see Wagner \& Witzel 1995 for a review). In the
source-intrinsic explanation, the flux variability on timescales
of less than one day would imply very small sizes of the emitting
regions and very high brightness temperatures (e.g. $10^{19}$K for
S4 0917+624, see Quirrenbach et al. 1989; Qian et al. 1991), seven
orders of magnitude greater than the inverse-Compton limit of $10^{12}$K
(Kellermann \& Pauliny-Toth 1969). These high brightness
temperatures can only be reconciled with the inverse-Compton
limit, assuming strong relativistic Doppler-boosting, with Doppler
factors larger than $\sim 100$. Such high Doppler factors are not
seen in VLBI. Alternatively, source extrinsic propagation effects,
e.g. the interstellar scintillation (ISS) caused by the
intervening line-of-sight material in our Galaxy, could be
responsible for the observed IDV of compact sources (Rickett et
al.1995; Rickett 2007). In order to better understand the
long-term behavior of IDV sources, we carried out an IDV
monitoring program for a number of IDV sources, including quasar
S4 0917+624 at the 6 cm band from August 2005 to January 2010 with the
Urumqi 25 m radio telescope in Xinjiang Astronomical Observatory,
China.

The quasar S4 0917+624 (OK 630, z=1.453) was one of the first
strong IDV sources found (Heeschen et al. 1987) with high
variability amplitudes of $\sim$10-15\% occurring on timescales of
$\leq$1 day. Jauncey \& Macquart (2001) and Rickett et al. (2001)
analyzed IDV data of S4 0917+624 and find a systematic variation
of the variability timescale over the year. This effect -- known as
the annual modulation of IDV -- is caused by the Earth orbiting
around the Sun and corresponding systematic variations of the
Earth velocity vector with respect to the stationary-assumed
scintillating pattern. For more examples of the annual modulation
of the IDV timescales, see Dennett-Thorpe \& de Bruyn (2000); Bignall et al. (2003, 2006); Gab\'anyi et
al. (2007); Marchili et al. (2012); Liu et al. (2012b, 2013a).

In September 1998 the previously strong IDV of S4 0917+624 changed
and suddenly became weaker (Kraus et al. 1999). This may be due to
a prolonged variability timescale, which in the ISS model for S4
0917+624 is expected to appear in September (Rickett et al. 2001).
The IDV started again in February 1999; however, from September 2000 to
2001 the IDV of S4 0917+624 then ceased completely (Fuhrmann et
al. 2002; Kraus et al. 2003). The intermittency of IDV in S4
0917+624 may be due to changes in the interstellar medium
properties and/or caused by an increased size of the
scintillating components (Narayan 1992; Walker 1998; Krichbaum et
al. 2002). Liu et al. (2013b) studied the total flux densities and
spectral index of S4 0917+624, and found that the source spectral
index became slightly steeper after the year 2000, implying that the
source may become less core-dominated after 2000. A correlation
between source spectral index and the ISS characteristics was
found in a large sample of 140 sources by Koay et al. (2011). IDV
sources may also be more variable intrinsically on longer
timescales than non-IDV sources (Peng et al. 2000). However, high-resolution VLBI data are required to study whether variations in
the source structure play a role in the variation of the IDV
pattern of S4 0917+624.

In section 2, we analyze our results from the Urumqi flux density
monitoring observations together with previous IDV data of S4
0917+624; in section 3, we study the long-term VLBI structural
variability of the source; in section 4, we discuss whether an overall
relationship between rapid flux density and structural variability
exists.

\section{Urumqi observations and data reduction}

We carried out IDV observations of S4 0917+624 approximately once a month at 4.8 GHz from August 2005 to January 2010 with the
Urumqi 25 m radio telescope, in dual polarization with the central
frequency of 4.8 GHz and bandwidth of 600 MHz. The typical system
temperature is 24~K in clear weather, and the antenna sensitivity
is $\sim$0.12 K/Jy.

The IDV observations were performed in cross-scans mode,
consisting of eight sub-scans in azimuth and elevation over the source
position. After initial calibration of the raw data, the intensity
profile of each sub-scan was fitted with a Gaussian function after
subtracting a baseline; then the fitted scans were averaged in
azimuth and elevation. After this, a correction for
residual pointing errors was made and the elevation and azimuth
scans were averaged together. In the next step an antenna gain-elevation correction was applied, including a correction for
airmass. The antenna gain-elevation correction was derived from
frequent observations of secondary calibrators observed during
each observing run. These calibrators were then also used to
correct the data for systematic time-dependent effects. Finally,
the raw amplitudes were converted to the absolute flux density
using the average scale of the primary calibrators, e.g. 3C48,
3C286, and NGC7027 (Baars et al. 1977; Ott et al. 1994).
The duration of each IDV monitoring session is about 3 days,
with typically one sample per hour for each target source
(where a sample is a set of eight sub-scans), from Table~\ref{tab1}.

For the data analysis, we use the statistical quantities, such as
the modulation index $m$ (a measure of the amplitude of the
variations); the variability amplitude $Y$, which is unbiased to
the systematic calibration uncertainty; and the reduced chi-squared $\chi_{r}^2$, to
describe the variability as shown below (see also Kraus et al.
2003),

\begin{equation}
m[\%] = 100\frac{\sigma_{s}}{<S>},
\end{equation}

\begin{equation}
Y[\%] = 3\sqrt{m^2-m_0^2},
\end{equation}

\begin{equation}
\chi_r^2 = \frac{1}{N-1}\sum(\frac{S_{i}-<S>}{\Delta S_{i}})^2,
\end{equation}

where $\sigma_{s}$ is the standard deviation of flux densities
from the mean flux density in a light curve, $<S>$ is the mean
flux density in the light curve, $m_{0}$ is the mean modulation
index of all secondary calibrators in the same observation, $N$ is
the number of measurements in the light curve, and $S_{i}$ and $\Delta
S_{i}$ denote individual flux density and its error.

\subsection{Result from Urumqi observations of S4 0917+624}

\begin{figure}
     \includegraphics[width=8.0cm]{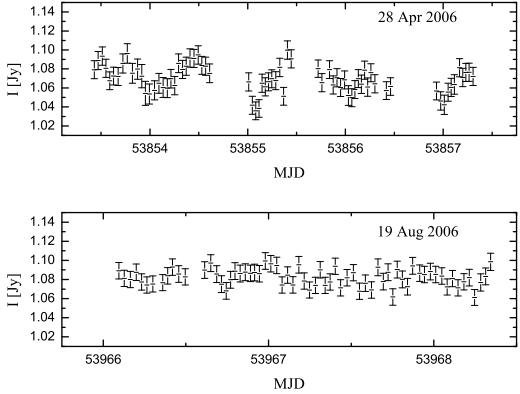}
     \caption{Upper panel: the weak IDV detected in the session of April
      2006; lower panel: non-IDV detected in the session of
      August 2006, see the definition of IDV identification in the text.}
      \label{fig1}
   \end{figure}

In order to ascertain whether or not IDV is present, we applied a
$\chi^2$ test to each data set. We adopt the criterion that a data
set with a probability of $\leq0.1\%$ of being constant is considered
to be variable, which is equivalent to the confidence level of
99.9\% to define the IDV. From this analysis, we find that the
source S4 0917+624 shows IDV in only 6 sessions, and does not show
IDV in 24 sessions in Table~\ref{tab1}. It may also be useful to
look at the ratio of the variability index $m$ to the mean
modulation index of calibrators $m_0$ (treated as 1-$\sigma$):
only two sessions show IDV at a 3-$\sigma$ level and six sessions
show IDV at greater than 2.5-$\sigma$. We therefore conclude that
the strong IDV of S4 0917+624, which was frequently observed before
the year 2000, was not present during our observations. As an example we show in Fig.~\ref{fig1} the light
curve in April 2006, which shows a weak IDV and the light curve in
August 2006, which does not show any significant variability.

To estimate the variability timescale from the six detected
weak IDV light curves (null probability $<$0.001 as defined
above), we applied the structure function $D(\tau)$ method
(Simonetti et al. 1985). In the first-order structure function, the shortest characteristic
variability timescale is located at the first plateau
where the structure function saturates. A structure function
$D(\tau)$ that monotonically increases with time lag can
be described by a power law. The shortest variability timescale is then defined by the intersection of this power-law fit
with a horizontal line, which fits the plateau (see Marchili et
al. 2012; Liu et al. 2012b). In the present paper, for consistency we used the Rickett et
al. (1995, 2001) definition of IDV timescale, which is defined at
half of the $D(\tau)$ saturation level, i.e. $\tau_{0.5}$, in Table~\ref{tab1} and
in the Appendix. Since the plateau or saturation level
sometimes is not well defined for the weak IDV, the estimated timescale
is only tentative for the Urumqi data in Table~\ref{tab1}.

\begin{table*}
         \caption[]{The results from the IDV observations of quasar S4 0917+624 with the Urumqi 25~m telescope at 4.8 GHz.}
         $$
         \begin{tabular}{cccccccccccc}


\hline
  \hline
    \noalign{\smallskip}
    1&2 &3 &4 & 5&6 &7 & 8&9 & 10 &11 &12\\
    Start day & dur & N & $<S>$ & $\sigma_{s}$  & $m$ & $m_{0}$ & $Y$ & $\tau_{0.5}$ & $\chi_r^2$ & $P_{null}$& $m/m_{0}$\\

       & [d] & & [Jy] & [Jy] & [\%] & [\%] & [\%] & [d] &&\\

\hline
  \noalign{\smallskip}

2005 Aug 15 &   2.8 &   52  & 0.899 &   0.013   &   1.48    &   0.34    &   4.32    &   0.26 &  9.30 &  5.30E-70&   4.4\\
2006 Apr 28 &   3.9 &   71  & 1.069 &   0.014   &   1.33    &   0.45    &   3.76    &   0.24 &  2.18 &  4.45E-08&   3.0\\
2006 Jun 10 &   3.2 &   88  & 1.090 &   0.014   &   1.29    &   0.50    &   3.58    &   0.24 &  1.80 &  7.32E-06&   2.6\\
2006 Jul 14&   4.0 &   94  & 1.083 &   0.014   &   1.29    &   0.60    &   3.42    &   -    &   1.51 &  1.11E-03&   2.2\\
2006 Aug 19 &   2.2 &   66  & 1.082 &   0.009   &   0.81    &   0.45    &   2.02    &   -   &   1.06 &  3.58E-01&   1.8\\
2006 Sep 23 &   4.9 &   141 & 1.093 &   0.010   &   0.96    &   0.45    &   2.54    &   - &  1.30 &  9.46E-03&   2.1\\
2006 Nov 17 &   4.7 &   133 & 1.092 &   0.012   &   1.08    &   0.47    &   2.91    &   - &     1.11 &  1.90E-01&   2.3\\
2007 Jan 25 &   2.3 &   66  & 1.132 &   0.009   &   0.82    &   0.41    &   2.14    &   -   &   0.94 &  6.09E-01&   2.0\\
2007 Feb 12 &   4.0 &   109 & 1.145 &   0.014   &   1.18    &   0.41    &   3.32    &   0.24 &  1.60 &  8.47E-05&   2.9\\
2007 Mar 24 &   2.8 &   71  & 1.163 &   0.009   &   0.76    &   0.46    &   1.82    &   - &     0.68 &  9.81E-01&   1.7\\
2007 Apr 20 &   3.6 &   78  & 1.211 &   0.017   &   1.36    &   0.62    &   3.65    &   0.18 &  1.62 &  4.80E-04&   2.2\\
2007 Jun 15 &   2.4 &   62  & 1.228 &   0.015   &   1.26    &   0.55    &   3.39    &   - &  1.64 &  1.23E-03&   2.3\\
2007 Jul 19 &   2.9 &   72  & 1.267 &   0.017   &   1.36    &   0.55    &   3.72    &   0.16 &  1.92 &  5.31E-06&   2.5\\
2007 Oct 13 &   3.0 &   59  & 1.261 &   0.007   &   0.59    &   0.34    &   1.42    &   - &     0.48 &  1.00E+00&   1.7\\
2007 Dec 21 &   3.2 &   76  & 1.285 &   0.009   &   0.71    &   0.32    &   1.91    &   -   &   0.67 &  9.89E-01&   2.2\\
2008 Feb 25 &   2.9 &   59  & 1.313 &   0.016   &   1.19    &   0.59    &   3.09    &   -&   1.50 &  7.88E-03&   2.0\\
2008 Mar 22 &   3.0 &   80  & 1.324 &   0.014   &   1.03    &   0.36    &   2.90    &   -   &   1.40 &  1.10E-02&   2.9\\
2008 Apr 22 &   3.1 &   74  & 1.347 &   0.012   &   0.87    &   0.47    &   2.19    &   -   &   0.87 &  7.85E-01&   1.9\\
2008 Jun 21 &   3.5 &   59  & 1.405 &   0.011   &   0.76    &   0.46    &   1.84    &   - &     0.86 &  7.75E-01&   1.7\\
2008 Jul 20 &   2.5 &   57  & 1.419 &   0.014   &   1.02    &   0.57    &   2.51    &   -   &   0.77 &  8.91E-01&   1.8\\
2008 Aug 20 &   5.0 &   70  & 1.442 &   0.015   &   1.06    &   0.60    &   2.63    &   - &     1.30 &  4.98E-02&   1.8\\
2008 Sep 13 &   3.4 &   88  & 1.467 &   0.012   &   0.83    &   0.35    &   2.27    &   -   &   0.82 &  8.90E-01&   2.4\\
2008 Nov 06 &   3.6 &   59  & 1.450 &   0.017   &   1.18    &   0.60    &   3.06    &   - &     0.94 &  6.11E-01&   2.0\\
2008 Dec 22 &   2.3 &   61  & 1.484 &   0.011   &   0.71    &   0.31    &   1.92    &   -   &   0.47 &  1.00E+00&   2.3\\
2009 May 13 &   2.7 &   51  & 1.530 &   0.021   &   1.39    &   0.58    &   3.80    &   -&   1.63 &  3.40E-03&   2.4\\
2009 Jun 25 &   2.6 &   53  & 1.510 &   0.020   &   1.32    &   0.57    &   3.55    &   - &  1.71 &  1.15E-03&   2.3\\
2009 Sep 22 &   5.5 &   135 & 1.533 &   0.014   &   0.95    &   0.55    &   2.30    &   - &     0.79 &  9.65E-01&   1.7\\
2009 Oct 09 &   2.3 &   61  & 1.528 &   0.012   &   0.78    &   0.38    &   2.06    &   - &     0.58 &  9.96E-01&   2.1\\
2009 Nov 22 &   3.8 &   78  & 1.523 &   0.019   &   1.24    &   0.72    &   3.05    &   -   &   0.99 &  7.43E-01&   1.7\\
2010 Jan 19 &   3.5 &   70  & 1.521 &   0.015   &   1.00    &   0.57    &   2.45    &   -   &   0.80 &  8.90E-01&   1.8\\

           \noalign{\smallskip}
            \hline
           \end{tabular}{}
         $$
         \label{tab1}
   \end{table*}

  \begin{figure}
     \includegraphics[width=9.5cm]{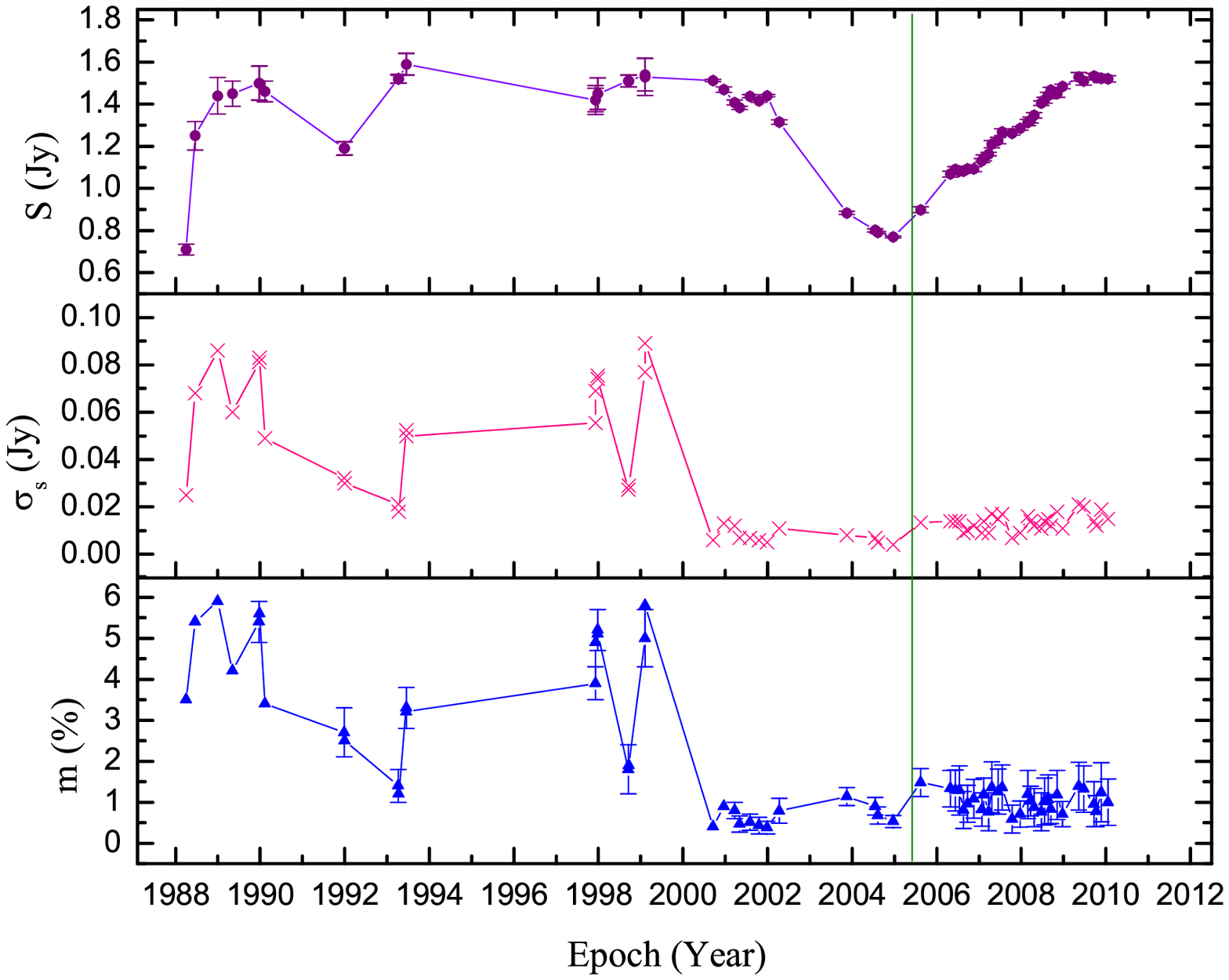}
     \caption{Upper panel: the total flux density of S4 0917+624 at $\sim$5 GHz in the IDV observations;
     middle panel: the rms flux density at $\sim$5 GHz; lower panel: the IDV modulation index at $\sim$5 GHz versus
time, the error on $m$ is the mean modulation index of
calibrators. Urumqi data from August 2005 to January 2010 are
shown to the right of the vertical line.
    }
      \label{fig2}
 \end{figure}

The results derived from the 30 observing sessions of S4 0917+624
are shown in Table~\ref{tab1}, the columns are: (1) observing
epoch; (2) the duration of observation; (3) the number of
effective data points; (4) the mean flux density; (5) the rms flux
density; (6) the modulation index $m$ of S4 0917+624; (7) the mean
modulation index of calibrators; (8) the relative variability
amplitude $Y$; (9) the IDV timescale with Rickett et al.
definition; (10) the reduced chi-square; (11) the null (non-variability) probability; (12) the ratio of the modulation index
to the mean modulation index of calibrators.

\subsection{Historical IDV characteristics of S4 0917+624}

In the Appendix we list the results of previous IDV observations for S4
0917+624 (1988-2004) taking from the literature. The
total flux density, rms flux density, and the modulation index of IDV in S4
0917+624 at $\sim$5 GHz, together with the Urumqi data from August
2005 to January 2010, are shown in Fig.~\ref{fig2}. It is obvious
that after the year 2000, the IDV of S4 0917+624 is very weak or
has ceased. Using the IDV data before 2000 (shown in the Appendix) we
obtain the following results: the total flux density shows no
correlation with the modulation index; the rms flux density shows
a correlation with the modulation index; and the IDV timescale
shows no correlation with either the total flux density or the
modulation index. The total flux density from the Urumqi
observations shows a monotonous increase over the 4.5 years, and
during this time the IDV of S4 0917+624 has been mostly quenched.
We confirm the lack of strong IDV after the year 2000. We note, however, that some weak IDV is present
in six epochs of the Urumqi observations (Table~\ref{tab1}) and
may also be in some epochs during 2003-2004 in the Effelsberg data (Bernhart
2010, see the Appendix). Before the year 2000 (see Fig.~\ref{fig2} and
the Appendix), the source frequently showed strong IDV, but we
note that times exist during which the IDV was also weak (e.g. in
April 1993 and in September 1998). This implies some intermittency of
the IDV of S4 0917+624 also appeared before the year 2000.

\section{Long-term VLBI structure changes of S4 0917+624}

The radio image of S4 0917+624 appears point-like on kpc scales, and
VLBI images at 5 GHz reveal a one-sided core-jet structure (Standke et
al. 1996; Gabuzda et al. 2000). A comparison between the IDV
strength and structural changes observed with VLBI may help to
explain and understand the substantially reduced IDV amplitudes after the year
2000. The IDV of S4 0917+624 was mainly observed at 5 GHz in the
past, unfortunately the source was not regularly monitored with
VLBI at 5 GHz. However, the source has been monitored with the VLBA at 15
GHz since 1995 (MOJAVE program, see Lister et al. 2009). The 15 GHz
VLBI data have a higher angular resolution than at 5 GHz, and are suitable
for the study of the structural evolution of the compact components in S4 0917+624.

\begin{figure}
  \centering
     \includegraphics[width=7cm]{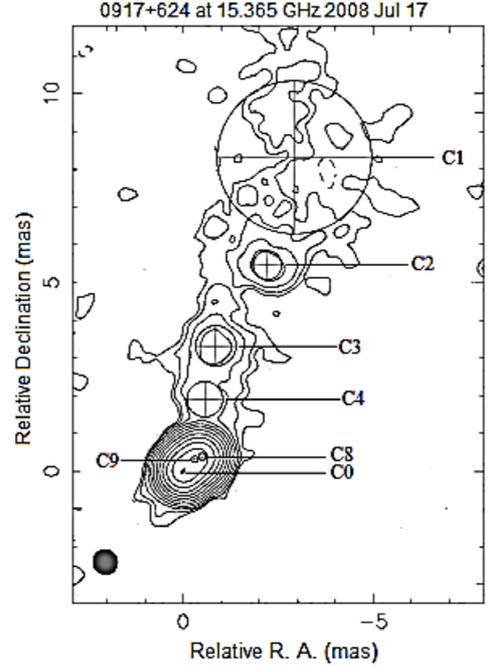}
     \caption{The VLBA map of S4 0917+624 at 15 GHz observed on 17 July 2008.
     the contours are the peak 0.81 Jy/beam times: -0.05, 0.05, 0.1, 0.2, 0.4, 0.8, 1.6, 3.2, 6.4, 12.8, 25.6,
51.2 (\%); the beam FWHM: 0.664 $\times$ 0.632 mas with the major
axis at $21.7^\circ$.}
      \label{fig3}
   \end{figure}

We downloaded the calibrated data (January 1995 to November 2010) of S4
0917+624 at 15 GHz from the MOJAVE database and fitted the source
structure using Gaussian components within the DIFMAP
package. Figure~\ref{fig3} shows an example of a VLBA image at 15
GHz. The jet features were fitted with several circular Gaussian
components, with the VLBI core (C0) registered at the southernend
of the jet-like brightness distribution. Each data set was
model-fitted several times to obtain a better error estimation.
The typical errors are $\sim$5\% for flux density and $\sim$0.05
mas for angular size and position of the fitted VLBI components,
respectively. For their cross-identification in the maps obtained
at different epochs, we used the VLBI core as the reference point
and assumed that it is stationary. With this method we identify the jet
components at different epochs through their distances from the
VLBI core, their flux densities, sizes, and relative position
angles. The resulting identification is similar to that described
in Lister et al. (2013).

\begin{figure}
   \centering
     \includegraphics[width=8.5cm]{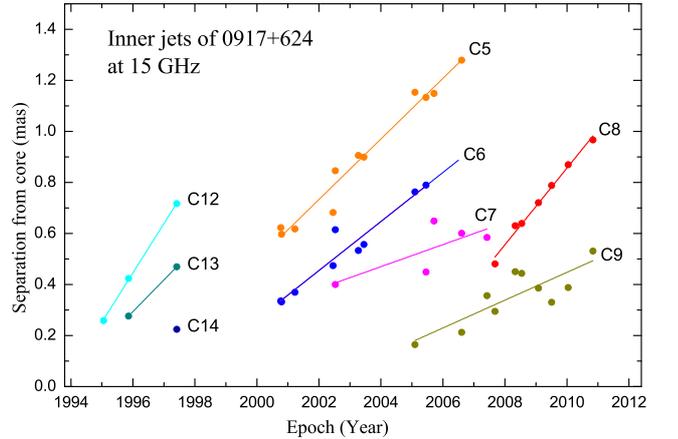}
     \caption{The separation of the model-fitted components relative to the core component against time. The solid
      lines are the linear fittings. We note that the identifications of components C7, C12, C13, and C14 are only tentative.}
      \label{fig4}
   \end{figure}

In Fig.~\ref{fig4}, we show the proper motion of the jet
components with respect to the stationary assumed VLBI core
component. The back-extrapolation of the motions of the jet
components C3, C4, C5, C6, C8, and C9 to zero-separation from the
VLBI core results in approximate ejection epochs around 1986.1,
1992.3, 1995.8, 1997.2, 2004.2, and 2001.8. In Table~\ref{tab2} we
summarize the kinematical parameters of the jet components at 15 GHz.
For the determination of the apparent speeds, we adopt
$\Lambda$CDM standard cosmology, with the following parameters:
$H_{0} = 71$ kms$^{-1}$ Mpc$^{-1}$, $\Omega_{M} = 0.27$, and
$\Omega_{\Lambda} =0.73$. Taking into account the opacity effect
(Sokolovsky et al. 2011; Pushkarev et al. 2012), we note that the
ejection time of a jet component may be delayed at 5 GHz when
compared to that at 15 GHz. Furthermore, it is found that there are helical
jet trajectories from the inner to outer jets, but the inner jets can
be fitted quasi-linearly (Bernhart 2010).

\begin{table}
         \caption[]{The fitted Gaussian components and derived properties from the VLBA images at 15 GHz.}
         $$
         \begin{tabular}{cccccc}


\hline
  \hline
    \noalign{\smallskip}
    1&2 &3 &4 & 5&6\\
ID & N & $T_{0}$& $\beta_{app}$& $<Sc>$ &$<Size>$\\

       & & & [c]&[Jy]&[mas]\\

\hline
  \noalign{\smallskip}

C0  &   22  &       &           &   0.507   &   0.10  \\
C1  &   13  &       &   11.8$\pm$3.3    &   0.016   &   3.32\\
C2  &   22  &       &   -0.7$\pm$0.3    &   0.024   &   1.31 \\
C3  &   21  &   1986.1  &   10.1$\pm$1.0    &   0.015   &   0.87\\
C4  &   17  &   1992.3  &   8.9$\pm$1.0     &   0.024   &   0.74\\
C5  &   11  &   1995.8  &   8.1$\pm$0.5     &   0.089   &   0.36\\
C6  &   9   &   1997.2  &   6.6$\pm$0.7     &   0.202   &   0.14\\
C7  &   5   &       &           &   0.064   &   0.13\\
C8  &   7   &   2004.2  &   10.2$\pm$0.5    &   0.387   &   0.22\\
C9  &   10  &   2001.8  &   3.7$\pm$0.9     &   0.256   &   0.18 \\
C10 &   1   &       &           &   0.004   &   0.76  \\
C11 &   3   &       &           &   0.030   &   0.14  \\
C12 &   3   &       &           &   0.378   &   0.25  \\
C13 &   2   &       &           &   0.304   &   0.06  \\
C14 &   1   &       &           &   0.403   &   0.13  \\

           \noalign{\smallskip}
            \hline
           \end{tabular}{}
         $$
         \label{tab2}
           \tablefoot{Columns are as follows: (1) component number (C0 denotes the core);
           (2) the number of epochs with which the component is identified; (3) the ejection time of component;
           (4) the apparent transverse velocity; (5) the mean flux density of component;
           (6) the mean size of component.}

   \end{table}

In our model-fittings of S4 0917+624, 14 components are registered
as listed in Table~\ref{tab2}, the core and inner-jet components are
mostly within 1 mas scale. We note that for some components (C7,
C10, C11, C12, C13, C14) the identification remains tentative because of the lack of data. The other components, however, are well
traced by their systematic proper motions. The component C14 may
be related to C5 or C7. The VLBI components that might affect the IDV
behavior of the source should be strong and compact (e.g. see Lee
et al. 2008). Therefore only the core and compact inner-jet
components will contribute substantially to the overall
scintillation pattern of the source. In the following, we will not
consider the diffuse/weak components C1, C2, C3, C4, C10, C11
(with mean flux $<0.05$ Jy in Table~\ref{tab2}) and component C7
(which is quite weak and not well identified).

The components C5 and C6 were ejected in 1995.8 and 1997.2
respectively at 15 GHz, and both are quite strong in flux density.
An emerging component can lead to an expansion of the apparent
core size, at a time when it is not yet well separated from the
core. An increased `core size' could reduce the IDV amplitude
(Walker 1998; Krichbaum et al. 2002). The weak IDV in September 1998
might be due to component C5 being ejected in 1995.8, if C5
was still blending with the core at 5 GHz. Afterwards from September
2000 to 2001, the IDV of S4 0917+624 ceased completely
(Fuhrmann et al. 2002; Kraus et al. 2003), which might have been caused
by C6 being ejected in 1997.2. The components C9 and C8 were
ejected in 2001.8 and 2004.2, both are strong in flux density. The
component C8 exhibits a higher apparent velocity of $\beta_{app} =
(10.2\pm0.5) c$ than component C9 of $\beta_{app} = (3.7\pm0.9)
c$. The components C9 and C8 might have partially reduced the IDV of S4
0917+624 after 2003. Pushkarev et al. (2012) estimated the VLBA
core shifts between 15.4 GHz and 8.1, 8.4, 12.1 GHz for the MOJAVE
sources. With their result for S4 0917+624 and the relation of core
shift versus frequency $r_{c} \propto \nu^{-1}$ (see Pushkarev et al.
2012), the core shift between 15 and 5 GHz can be estimated, and
then with the proper motion listed in Table~\ref{tab2} we estimate
the time delay for the components C5, C6, C8, and C9 are of
1.4$\pm$0.2, 1.7$\pm$0.4, 1.1$\pm$0.1, and 3.1$\pm$1.5 years,
respectively, between 15 and 5 GHz.

\begin{figure}
\centering
    \includegraphics[width=9cm,height=12cm]{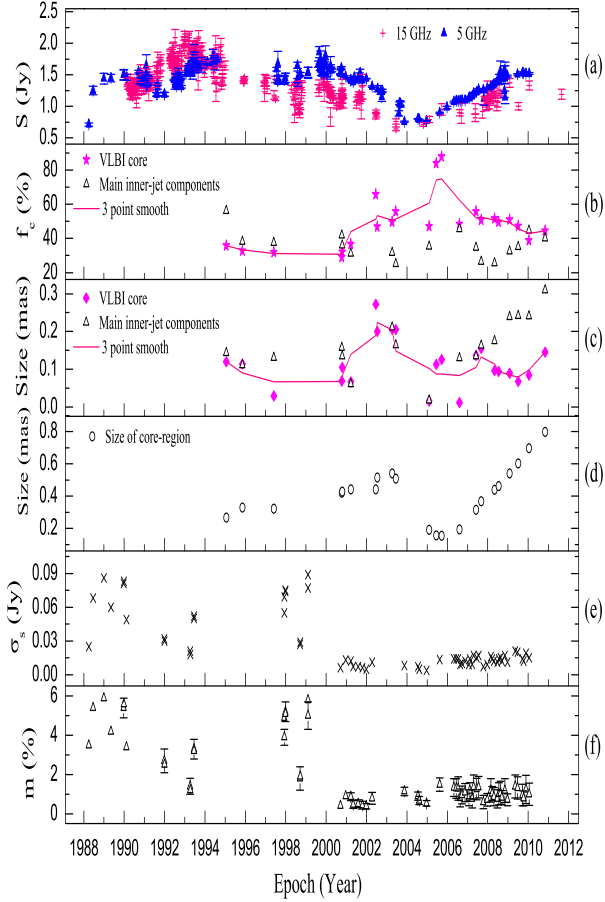}
     \caption{From top to bottom: (a) the total flux density at 14.5 GHz (from the UMRAO database)
     and at 4.8 GHz (from the UMRAO and the IDV observations); (b) the flux ratio ($f_{c}$)
     of the VLBA core or inner-jet component to total flux density of S4
     0917+624; (c) angular size of the VLBA core or inner-jet components at 15
     GHz; (d) size of the broad `core region' fitted with a single Gaussian model; (e) the rms flux density at 5 GHz versus time;
     (f) the IDV modulation index at 5 GHz versus time.
     }
     \label{fig5}
   \end{figure}

The ratio ($f_{c}$) of the flux density of VLBI component to source total flux density
shows that the 15 GHz VLBI core has flux ratios
of $f_{c}\sim $28\%-88\%, and nearly half of the inner-jet components within 1 mas from the core
have $f_{c}> 25\%$. In Fig.~\ref{fig5}b the flux ratios for the VLBI
core and inner-jet components with $f_{c}> 25\%$ are plotted. We note that the
flux ratio of the VLBI core is generally higher after the year 2000.

In Fig.~\ref{fig5}c, the angular sizes of the VLBI core
and main inner-jet components versus time are illustrated. The angular size is
defined as the FWHM of the fitted circular Gaussian (in some
epochs the core was better fitted by an elliptical Gaussian, for
which we use the geometrical mean of the major and minor FWHM).
Figure~\ref{fig5}c shows that the core size significantly
increased in 2002-2003, but before and after this period the core
sizes were similarly low. The sizes of the inner-jet components
have increased since 2007 as a result of the expansion of components C8 and
C9. The increased size of the VLBI core in 2002-2003 may have led
to a partially quenched IDV of S4 0917+624. However, the sizes of
the VLBI core and inner-jet components are relatively small in
2005-2006 and comparable to that before year 2000, and no strong
IDV has been observed during this period. It is also helpful to
fit the broad `core region' (see Fig.~\ref{fig3}) with a single
elliptical Gaussian model rather than with multiple components;
the resulting size evolution in Fig.~\ref{fig5}d is similar to the
trend seen for the inner-jet components.

\begin{figure}
\centering
    \includegraphics[width=8cm]{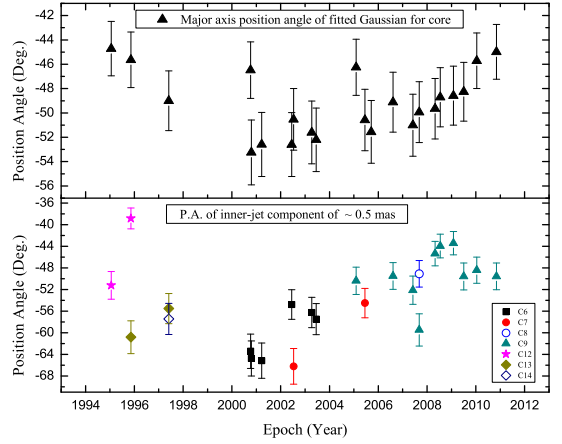}
     \caption{Upper panel: the position angle of the 15 GHz VLBA core versus time; lower panel:
     the position angles of the inner-jet components versus time.}
     \label{fig6}
   \end{figure}

\section{Discussion}

The VLBI images of S4 0917+624 at 5 GHz show a $\sim$60-70\% of
total flux density in the core region ($<$ 1 mas) at 5 GHz
(Standke et al. 1996; Gabuzda et al. 2000). The core region is
resolved into a 15 GHz core and inner jets (see Fig.~\ref{fig3}),
with a lower core fraction (Fig.~\ref{fig5}b) at 15 GHz than
that of $\sim$60-70\% at 5 GHz. Rickett et al. (1995) made a
detailed and sophisticated analysis of S4 0917+624 during its
phase of large-amplitude IDV. The results of the detailed
modeling with the refractive ISS (RISS) by Rickett et al. (1995)
show that the scintillating component of S4 0917+624 at 5 GHz has
an angular size of $\sim$ 70 $\mu as$ with flux fraction of $\sim$ 44\%,
under the brightness temperature of $\sim 6\times 10^{12}$K.
Assuming the scintillating component resides in the VLBI core, for
the modulation index $m_{s}$ of the scintillating component with the
flux fraction $f_{s}$, and the modulation index of total flux
density $m$, we have

\begin{equation}
m = f_{s} m_{s}(\theta_{r}/\theta_{s})^{7/6},
\end{equation}

where $\theta_{r}$ is the angular size of refractive scattering
disk and $\theta_{s}$ is the size of scintillating source
component (see e.g. Rickett et al. 1995; Walker 1998).

For a point source (i.e. $\theta_{s}=\theta_{r}$), by taking the
value $m\sim 5.5\%$ in the episode strong IDV phase before the
year 2000 (in the Appendix) and $f_{s}=44\%$, we have $m_{s}\sim
13\%$. This is much lower than the value of $m_{p}\sim$ 50\% for a
point source estimated by Rickett et al. (1995), implying
that the scintillating component of S4 0917+624 is not a point source, i.e. the
size of the scattering disk is $<70 \mu as$ of the scintillating
component. Such small angular size was not resolved in the 5
GHz VLBI observations (Standke et al. 1996; Gabuzda et al. 2000),
but from the deconvolved core size in the model fitting on the 15 GHz VLBA data,
a higher resolution of $\sim 100 \mu as$ at 15 GHz
was estimated in some epochs (Fig.~\ref{fig5}c) for S4 0917+624.
It is possible that changes in a very high brightness ($\leq 100 \mu as$)
component embedded in the core region could be related to
changes in the ISS. Such high
radio brightness of AGN cores has been reported to be significantly
above the known inverse-Compton limit in the Radioastron space
VLBI observations (Kovalev 2014).

Liu et al. (2013b) noted that the spectral index between 5 and 15 GHz (defined
as $S \propto \nu^{+\alpha}$) became slightly steeper, being in the range -0.1 to 0.2 before the year 2000,
and -0.3 to -0.1 after 2000. This may have been caused by the new jet
components C6, C8, and C9 evolved after 2000. The jets usually have
steeper spectral indices. In Fig.~\ref{fig5}b, the core
fraction at 15 GHz increased after the year 2000, implying that
the 15 GHz VLBI core may blend with inner-jet components that
were not resolved with the 15 GHz VLBI yet. The VLBI core size
can be enlarged owing to the unresolved inner jet, and could
partially quench the IDV of S4 0917+624, e.g. for the VLBI core
size increased in 2002-2003 (Fig.~\ref{fig5}c). However in 2005-2006
the VLBI core size is comparable to that before 2000, and it is
difficult to explain why there was no strong IDV in this period.
Hovatta et al. (2014) investigated the spectral indices of the
MOJAVE sources observed at 8.1, 8.4, 12.1, and 15.4 GHz with the
VLBA, in which the VLBA core of S4 0917+624 on 9 August 2006 showed
very flat spectral index (0.04) as well as three inner jets
($\leq$ 0.5 mas from the core) with flat spectral indices
$|\alpha|<$ 0.1. This flat spectrum core would be able to
scintillate, but no IDV has been found from the Urumqi observation
in August 2006 (see Table~\ref{tab1}).

The position angle changes of source components may also affect
the scintillation (Liu et al. 2012b). We find that the position
angles (PA) of the 15 GHz VLBI core and inner jets changed a
few degrees to the west after the year 2000
(Fig.~\ref{fig6}), and this could induce an apparently larger core
size in projection. This effect may be significant if the scintillating component is
extremely anisotropic. In January 2010, the PA
of the 15 GHz VLBI core returned back to $\sim$ -46$^{\circ}$, but
no strong IDV was detected, suggesting that the quenching effect
due to the small PA changes is not significant for this source.

The intra-day flux variations of S4 0917+624 showed no time delay
from different radio bands (Rickett et al. 1995). Long-term variations of AGN
have shown time delays of a few tens of days between different radio bands (Fuhrmann et al. 2014).
With the radio light curves of S4 0917+624 at 4.8 GHz and 14.5 GHz from the UMRAO
data, as well as the data from the 5 GHz IDV observations in
Table~\ref{tab1} and the Appendix (see Fig.~\ref{fig5}a), we
estimated an upper limit of $\sim$ 1 year for the time delay
between the 15 and 5 GHz with the cross correlation analysis. This
upper limit is close to but lower than that from the core shift
estimation in the previous section. And we should note that both the VLBA
data at 15 GHz and the IDV data at 5 GHz are relatively sparse and
do not exactly match each other in time. This might prevent us from
detailed comparison of the VLBI structure changes and the IDV
evolution; for instance, there are no IDV data around the VLBA
observations during 1995.0-1997.5. After the year 2000, however,
it is evident that the IDV characteristics in Fig.~\ref{fig5}e and
Fig.~\ref{fig5}f have no correlation with the VLBI structural
changes in Fig.~\ref{fig5}b to Fig.~\ref{fig5}d, although there
are weak IDV in three epochs from August 2005 to June 2006 (see
Table~\ref{tab1}) when the 15 GHz VLBA core size is relatively
small (see Fig.~\ref{fig5}c and also the images in the MOJAVE
webpage\footnote{
www.physics.purdue.edu/astro/MOJAVE/sourcepages/}).

The disappearance of strong IDV in S4 0917+624 after the year 2000
cannot be fully explained with the quenching mechanism via changes
of core size in the source, as discussed above. Bernhart et al.
(2006) studied the VLBI kinematics of S4 0917+624 with data during
2000-2007 at 5, 15, and 22 GHz, and also did not find clear
correlation between the IDV properties and VLBI structure. The
disappearance is likely caused by a change of ISM scattering properties, e.g. with the passage of scattering material out
of the line of sight to the quasar, as discussed for the fast IDV
source J1819+3845, which also ceased at some time in the period
between June 2006 and February 2007 (de Bruyn \& Macquart 2014), and
for the intermittent IDV source PKS 0405$-$385 (Kedziora-Chudczer
2006).

To identify the possible scattering materials located in the
foreground of S4 0917+624, which has the distance $\sim$ 200 pc
estimated with the shortest timescale in the Appendix by Rickett
et al. (1995, 2001), we have compared the source position with
existing data archive of various ISM materials (e.g. radio loops
and dust emission). The source lies on the border of Loop-III
(Berkhuijsen et al. 1971) about 150-200 pc from us, as well as on
the local enhancement of IRAS dust emission, which may be formed by
interactions of radio loops with the ambient medium or with other
structures. The consequence of such a collision is a dense turbulent
local structure, which is responsible for the required enhanced
scattering material producing ISS (Linsky et al. 2008). Ramachandran et al. (2006) found that the small-scale
properties of the intervening plasma as it drifts through the
sight line is evident with dispersion measure variations of pulsar
timing, which causes the refractive scintillation timescales from
days to months. However, further studies on the interstellar weather toward
S4 0917+624 are required, e.g. with multi-frequency single dish
and VLBI polarimetry to measure Faraday rotation and possibly
detect variations of the rotation measure in the foreground screen
(e.g. see Gab\'anyi et al. 2009), and with spectral line observations
(Malamut et al. 2014) and the observation of pulsars near the
line of sight to detect the ISM intermittency, which could be
related to the changing IDV pattern.

\section{Summary and conclusion}

We presented the result of S4 0917+624 observed with the Urumqi
25~m telescope from August 2005 to January 2010. The quasar exhibits almost no IDV in our observations, which is in
contrast with the previous pronounced IDV before year 2000.
We also analyzed the long-term VLBI structural variability by using Gaussian
model-fitting of the 15 GHz VLBA data, and obtained the flux densities and the
deconvolved sizes of core and inner-jet components of the
source. The source shows ejection of several jet components that
are suspected to have partially reduced the IDV amplitude of S4
0917+624. However, during 2005-2006, the VLBI core size was
comparable to the size before the year 2000, but no strong IDV
was detected in the period. We studied the properties such as core fraction, angular size,
spectral index, and brightness temperature of VLBI core for S4 0917+624, as well as the time delay between
5 and 15 GHz variations, and compared these values with
the IDV properties of S4 0917+624. With the current data, the vanishing strong IDV in S4
0917+624 after the year 2000 cannot be explained via the quenching effect by
changes in the observable VLBI structure. However, it may be caused
by changes in the interstellar medium, i.e. by interstellar
weather, which induces changes in the scintillation pattern on
timescales of several years. The refractive scattering properties for the strong IDV phase of S4 0917+624 before
the year 2000 have been discussed. Further coordinated multi-frequency
observations will be required to distinguish between the effect of source-intrinsic
variability and changing properties of the
interstellar medium.

\begin{acknowledgements}
We acknowledge the referee for valuable comments that have
improved the paper a lot. M.L.G. thanks M. Lister and G.-Y Zhao for
helpful comments on the model fitting method to the VLBA data.
This work is supported by the National Natural Science Foundation
of China (No.11273050); the 973 Program 2015CB857100; Key
Laboratory of Radio Astronomy, Chinese Academy of Sciences; and
the program of the Light in China's Western Region (Grant No.
YBXM-2014-02). N.M. is funded by an ASI fellowship under contract
number I/005/11/0. This research has made use of data from the
MOJAVE database that is maintained by the MOJAVE team (Lister et
al., 2009). This research has made use of data from the University
of Michigan Radio Astronomy Observatory which has been supported
by the University of Michigan and by a series of grants from the
National Science Foundation, most recently AST-0607523.
\end{acknowledgements}

\onecolumn
\begin{appendix}
\section{Collected historical IDV results of S4 0917+624 at $\sim$5 GHz.}
\centering
         \begin{tabular}{ccccccccccc}

\hline
  \hline
    \noalign{\smallskip}
    1&2 &3 &4 & 5&6&7&8&9&10&11\\

         Date & dur & N & $<S>$ & $\sigma_{s}$  & $m$  & $m_{0}$ & $Y$    & $\tau_{0.5}$($type$) & $\chi_r^2$ &ref \\

              & [d] &   & [Jy]    & [Jy]         &[\%]& [\%]    & [\%] & [d]        &             &\\

\hline
  \noalign{\smallskip}

1988 Apr 03 &   4.4     &       &   0.71    &   0.025   &   3.5 &       &       &   0.22$\pm$0.10   &       &   1   \\
1988 Jun 17 &   4.6     &       &   1.25    &   0.068   &   5.4 &       &       &   0.33$\pm$0.15   &       &   1   \\
1988 Dec 30 &   6.1     &       &   1.44    &   0.086   &   5.9 &       &       &   0.28$\pm$0.12   &       &   1   \\
1989 May 06 &   5.3     &       &   1.45    &   0.060   &   4.2 &       &       &   0.13$\pm$0.05   &       &   1   \\
1989 Dec 26 &   5.9     &       &   1.50    &   0.083   &   5.6 &       &       &   0.34$\pm$0.10   &       &   1   \\
1990 Feb 11 &   25.1    &       &   1.46    &   0.049   &   3.4 &       &       &   0.17$\pm$0.02   &       &   1   \\
1991 Dec 31 &   6.9     &       &   1.19    &   0.030   &   2.5 &       &       &   0.29$\pm$0.14   &       &   1   \\
1993 Apr 12 &   2.3     &       &   1.52    &   0.018   &   1.2 &       &   &   $\geq$0.13&       &   1   \\
1993 Jun 19 &   2.6     &       &   1.59    &   0.050   &   3.2 &       &   &   $\geq$0.15&       &   1   \\
1997 Dec 06 &   2.7     &       &   1.42    &   0.069   &   4.9 &       &   &   $\geq$0.55&       &   1   \\
1997 Dec 28 &   5.4     &       &   1.45    &   0.074   &   5.1 &       &   &   0.28$\pm$0.14   &       &   1   \\
1998 Sep 19 &   5.0     &       &   1.51    &$\geq$0.029 &$\geq$1.9&  &&  $\geq$1.33 &       &   1   \\
1999 Feb 09 &   2.2     &       &   1.53    &   0.089   &   5.8 &       &   &$\geq$0.37    &       &   1   \\
1989 Dec 22 &   6   &   111 &   1.50    &   &   5.4 &   0.5 &   16.2    &   II  &   104.0   &   2   \\
1991 Dec 27 &   7   &   62  &   1.19    &   &   2.7 &   0.6 &   7.7 &   II  &   20.4        &2  \\
1993 Apr 10 &   3   &   196 &   1.52    &   &   1.4 &   0.4 &   4.2 &   I   &   9.9     &2  \\
1993 Jun 18 &   2   &   186 &   1.59    &   &   3.3 &   0.5 &   9.7 &   II  &   42.8    &2  \\
1997 Dec 05 &   3   &   44  &   1.42    &       &   3.9 &   0.4 &   11.7    &   II  &   99.4    &2  \\
1997 Dec 25 &   6   &   43  &   1.45    &   &   5.2 &   0.5 &   15.4    &   II  &   123.6   &2  \\
1998 Sep 17 &   5   &   91  &   1.51    &   &   1.8 &   0.6 &   5.1 &   I   &   11.7        &2  \\
1999 Feb 08 &   6   &   18  &   1.54    &   &   5.0     &   0.7 &   14.9    &   II  &   54.8&2  \\
\hline
2000 Sep 16 &       &       &   1.513   &   0.006   &   0.4 &       &       &   &           &3  \\
2000 Dec 17 &       &       &   1.470   &   0.013   &   0.9 &       &       &   &           &3  \\
2001 Mar 24 &       &   57  &   1.419   &   0.012   &   0.87    &   0.20    &   2.53    &   &   2.803       &4  \\
2001 May 04 &       &   65  &   1.383   &   0.007   &   0.47    &   0.20    &   1.29    &   &   3.598       &4  \\
2001 Aug 03 &       &   138 &   1.438   &   0.007   &   0.51    &   0.20    &   1.41    &   &   4.108       &4  \\
2001 Oct 20 &       &   82  &   1.416   &   0.006   &   0.43    &   0.20    &   1.14    &   &   3.689       &4  \\
2001 Dec 26 &       &   38  &   1.440   &   0.005   &   0.38    &   0.15    &   1.04    &   &   3.220       &4  \\
2002 Apr 12 &       &   15  &   1.315   &   0.011   &   0.79    &   0.30    &   2.23    &   &   7.674       &4  \\
2003 Nov 14 &       &   47  &   0.884   &   0.008   &   1.14    &   0.22    &   3.37    &   &   17.508      &4  \\
2004 Jul 16 &       &   46  &   0.801   &   0.007   &   0.90    &   0.21    &   2.64    &   &   12.471      &4  \\
2004 Aug 12 &       &   112 &   0.791   &   0.005   &   0.68    &   0.21    &   1.93    &   &   6.652       &4  \\
2004 Dec 19 &       &   32  &   0.769   &   0.004   &   0.53    &   0.15    &   1.52    &   &   5.984       &4  \\

       \noalign{\smallskip}

            \hline
           \end{tabular}{}
           \tablefoot{Columns are as follows: (1) observing epoch, (2) the duration of observation, (3)
           the effective number of data points, (4) and (5) the mean flux density and the rms flux density,
           (6) the modulation index, (7) the mean modulation index of calibrators, (8) the relative variability amplitude of IDV,
           (9) the characteristic timescale in days (variability types: type II means an IDV timescale $<$ 2 days
           while type I means an IDV timescale $>$ 2 days), (10) the reduced chi-square,
            (11) the references: 1 Rickett et al. (2001), 2 Kraus et al. (2003),
           3 Fuhrmann et al. (2002), 4 Bernhart (2010).}

   \end{appendix}

   \end{document}